\documentclass[
  journal=largetwo,
  manuscript=research-paper,
  year=????,
  volume=??,
]{cup-journal}

\usepackage{amsmath}
\usepackage[nopatch]{microtype}
\usepackage{booktabs}

\title{4XMM~J182531.5$-$144036: A new persistent Be/X-ray binary found within the \emph{XMM-Newton} serendipitous survey}

\author{A.B. Mason}
\affiliation{School of Physical Sciences, The Open University, Walton Hall, Milton Keynes MK7 6AA, UK}
\alsoaffiliation{Finnish Centre for Astronomy with ESO (FINCA), University of Turku, V\"{a}is\"{a}l\"{a}ntie 20, FI-21500 Piikki\"{o}, Finland}

\author{A.J. Norton}
\affiliation{School of Physical Sciences, The Open University, Walton Hall, Milton Keynes MK7 6AA, UK}

\author{J.S. Clark$^\dag$}
\affiliation{School of Physical Sciences, The Open University, Walton Hall, Milton Keynes MK7 6AA, UK}

\author{S.A. Farrell}
\affiliation{Department of Physics \& Astronomy, University of Leicester, Leicester LE1 7RH, UK}

\author{A.J. Gosling}
\affiliation{Astrophysics, Department of Physics, University of Oxford, Denys Wilkinson Building, Keble Road, Oxford OX1 3RH, UK}

\email[A.J. Norton]{andrew.norton@open.ac.uk \\
$^\dag$ In memoriam: This paper is dedicated to the memory of our late colleague Simon Clark who inspired us all to try to understand the stars.}


\begin{document}

\begin{abstract}
We aim to investigate the nature of time-variable X-ray sources detected in the {\it XMM-Newton} serendipitous survey. The X-ray light curves of objects in the {\it XMM-Newton} serendipitous survey were searched for variability and coincident serendipitous sources observed by {\it Chandra} were also investigated. Subsequent infrared spectroscopy of the counterparts to the X-ray objects that were identified using UKIDSS was carried out using {\it ISAAC} on the VLT. We found that the object 4XMM~J182531.5--144036 detected in the {\it XMM-Newton} serendipitous survey in April 2008  was also detected by {\it Chandra} as CXOU~J182531.4--144036 in July 2004. Both observations reveal a hard X-ray source displaying a coherent X-ray pulsation at a period of 781~s. The source position is coincident with a $K=14$ mag infrared object whose spectrum exhibits strong HeI and Br$\gamma$ emission lines and an infrared excess above that of early B-type dwarf or giant stars. We conclude that 4XMM~J182531.5--144036 is a Be/X-ray binary pulsar exhibiting persistent X-ray emission and is likely in a long period, low eccentricity orbit, similar to X Per.
\end{abstract}

\maketitle

\section{Introduction}
\label{intro}
Be/X-ray binaries (BeXRBs) comprise the largest subset of the High Mass X-ray Binaries \citep{HMXRBcat}. They consist of an X-ray emitting neutron star pulsar in an (often eccentric) orbit around a Be star, i.e. a non-supergiant B-type star that has shown hydrogen lines in emission. Be stars also show an excess of infrared emission, compared to a star of the same spectral type. The emission lines and infrared excess are attributed to an equatorial circumstellar disc, due to material expelled from the rapidly rotating Be star. Some BeXRBs exhibit persistent X-ray emission whilst others only show such emission close to periastron, when the neutron star passes through the Be star circumstellar disc \citep{Reig11}.

Here we present the discovery of a new BeXRB, 4XMM J182531.5$-$144036, identified from an investigation of hard X-ray sources in the {\it XMM-Newton} serendipitous survey. The source was also detected by {\it Chandra} so we investigate those archival observations too. In Section  we introduce the X-ray observations and discuss the reduction of both data sets. The timing and spectral analysis of the X-ray source is then described within Sections 3 and 4 respectively. Section 5 examines near infrared observations and the identification of the counterpart to this system and Section 6 presents a discussion of our results.

\vspace*{-5mm}

\section{X-ray observations}
\label{X_ray_obs}

\subsection{XMM-Newton}

The field of 4XMM~J182531.5$-$144036 was observed on the 10th April, 2008 by {\it XMM-Newton} \citep{jansen01} for $33.5$~ks starting at MJD 54566.76197. The target was the Wolf-Rayet star WR~115 and the source 4XMM~J182531.5$-$144036 is located close to the edge of the field of view (see Figure 1, left). The observation was performed using all three EPIC cameras in imaging mode; the MOS1, MOS2 and PN cameras were employed in full frame mode using a medium filter. Additionally the Optical Monitor (OM) instrument performed optical/UV observations with the UVM2 filter.

\begin{figure*}[ht]
\includegraphics[width=18cm]{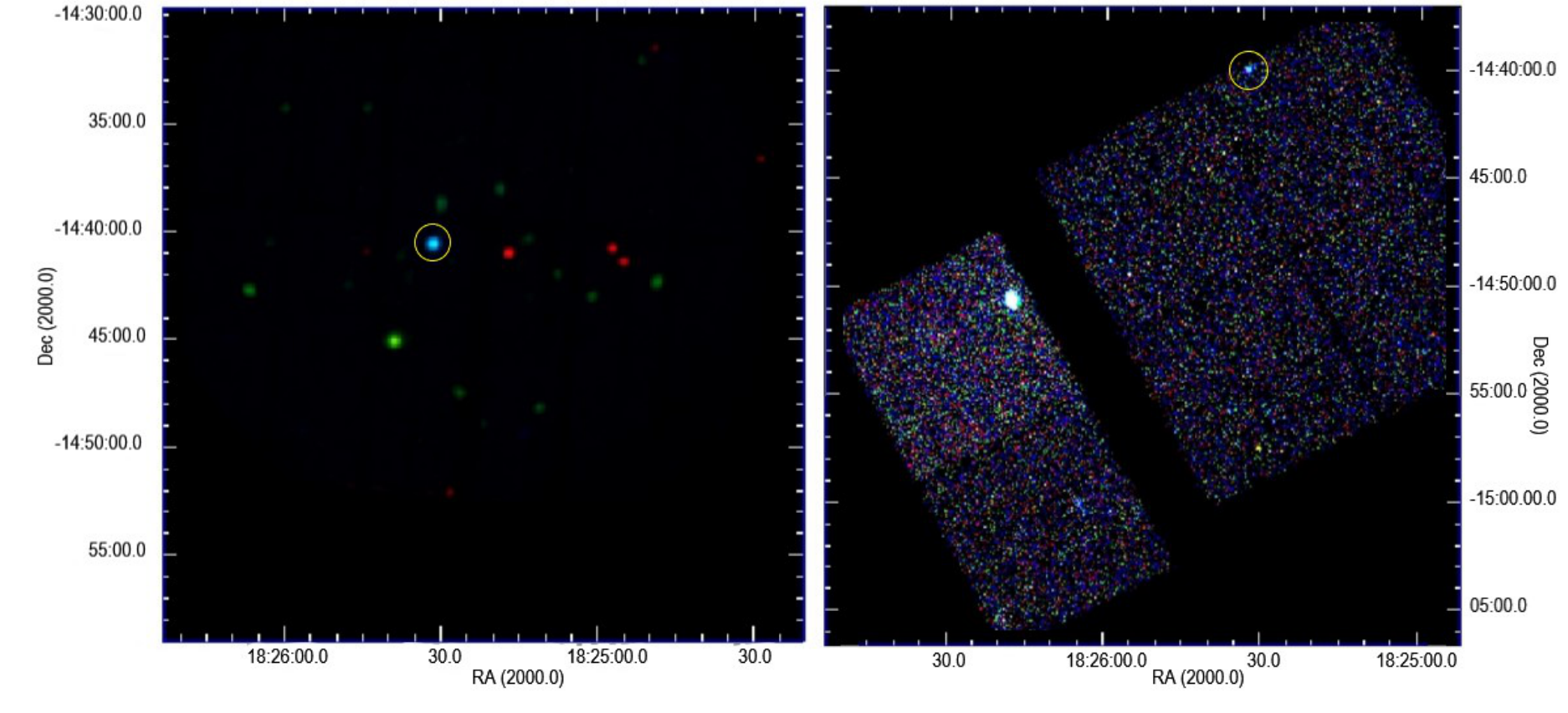}
\caption{Left: Merged \textit{XMM-Newton} EPIC RGB mosaic image (Red: 0.2--1 keV, Green: 1--2 keV, Blue: 2--10 keV). Gaussian smoothing with a kernel radius of 3 has been applied to the image. 4XMM~J182531.5$-$144036 is the ringed light blue source near the centre. Right: \textit{Chandra} ACIS-I merged RGB mosaic image (Red: 0.5--1.2 keV, Green: 1.2--2.0 keV and Blue: 2.0--7.0 keV. CXOU~J182531.4$-$144036 is the ringed source at the top of the right hand ACIS-I detector. Featured prominently on the left-hand side of the detector is the HMXB V479~Sct/LS~5039 \citep{hadasch12}.}
	\label{three_colour}
\end{figure*}

4XMM~J182531.5$-$144036 was initially detected by the {\it XMM-Newton} pipeline \citep{watson09} at RA: $18^{\text{h}}$~$25^{\text{m}}$~$31.5^{\text{s}}$ and Dec: $-14^{\circ}$~$40^{\prime}$~$36.6^{\prime\prime}$ (J2000) with a positional uncertainty of $1.1^{\prime\prime}$ ($3\sigma$). For this observation the {\it 2MASS} catalogue was used to correct the source position, providing a reduced pointing error of $0.35^{\prime\prime}$. No source was detected in the OM mosaiced image within the EPIC error circle.

Observation Data Files (ODFs) were processed using the {\it XMM-Newton} Science Analysis System (SAS) v21.0 (\url{http://xmm.esa.int/sas/}). Event lists for each PN and MOS camera were generated using the {\it epproc} or {\it emproc} script respectively, each employing the most recent calibration files, and using default parameters. Bad events caused by cosmic rays, edge effects and bad rows were identified and discarded. Single event light curves were constructed for all three EPIC cameras using energies greater than 10~keV to detect any episodes of soft proton flaring that may have occurred during the observation. For each camera a Good Time Interval (GTI) file was created to filter out these flaring episodes, with a background threshold of 1~count~s$^{-1}$ for the PN and 0.3~count~s$^{-1}$ for the MOS detectors. This resulted in net exposure times after flare removal of 28.0/32.8/32.9~ks for the PN/MOS1/MOS2 cameras respectively.

Source events were extracted for all three detectors in the energy range $0.2 - 10$~keV using an extraction region of radius $44^{\prime\prime}$ surrounding 4XMM~J182531.5$-$144036. Background events were also extracted from an identical sized circular region (free of any contaminating sources) and located at the same off-axis position. For the EPIC-PN spectrum and light curve we selected single pixel and double pixel events; for EPIC-MOS we used single to quadruple events. The source had EPIC PN
/MOS1/MOS2 count rates of $0.177 \pm 0.003$ / $0.056 \pm 0.002$ / $0.060 \pm 0.002$~counts~s$^{-1}$ respectively (background subtracted and corrected for vignetting and the point spread function). Images in three distinct energy bands, 0.2 -- 1 keV (Red), 1 -- 2 keV (Green) and 2 -- 10 keV (Blue) were extracted from the data and merged into a RGB mosaic (Figure 1) to display the spectral colours. The size of the 2-d bins used in making the images was 100 pixels per bin in both $x$ and $y$ direction (about 5 arcsec).

Source and background light curves with the maximum time resolution of 73~ms for the PN and 2.6~s for MOS1 and MOS2 were then created. To improve statistics we re-binned the PN light curve to a resolution of 2.6~s and combined this with the MOS1 and MOS2 light curves into a single light curve.  The GTIs for each camera largely overlap with each other, and all GTIs for each camera were used in the subsequent analysis. After background subtraction the resultant light curve was barycentrically corrected using the task {\it barycen}. Background and source spectra were then extracted for each EPIC camera, together with response and ancillary response files. The spectra were not re-binned, and the full spectral resolution was retained for the subsequent spectral fitting.

\subsection{Chandra}
\label{chandra_obs}

The {\it Chandra} source designated CXOU J182531.4$-$144036 was detected during an observation of the supernova remnant G16.85$-$1.05 (PI: Garmire) on the 11th July, 2004 for $\sim18$~ks starting at MJD~53197.51016. The observations employed the Advanced CCD Imaging Spectrometer (ACIS-I) \citep{garmire03b} in wide-field imaging mode utilising chips I0~--~I3 plus S2 and S3 in faint data mode. The data was reduced using {\it CIAO} v4.5 and the latest calibration files at that time  (calDB 4.5.7). Initially we reprocessed the default level 2 files using the {\it chandra$\_$repro} script to apply the current calibration set. All processing of the data was carried out with reference to the list of current Chandra data reduction threads (\url{http://cxc.harvard.edu/ciao/threads/}).

Selecting a circular background region with a radius of 42$^{\prime\prime}$ from the Level 2 event file, we created a light curve in the energy range 0.5 -- 7.0 keV to ascertain the degree of flaring. No periods of flaring were detected during the full exposure time. A three colour image of field was then constructed (see Figure 1, right). To convert the counts detected into fluxes, we then constructed an exposure map in the energy range 0.3 -- 8 keV with a centre-band energy of 2.3~keV.

To extract events in the energy range 0.2 -- 8 keV from CXOU J182531.4$-$144036 we used the {\it ChaRT} and {\it MARX} programs to calculate the radius given by 90$\%$ of the PSF at an energy of 4.5~keV. The calculated PSF is heavily dependent upon the off-axis and azimuthal angles. As the source is close to the edge of the detector, this yielded an off-axis angle $\theta = 8.3^{\prime}$ and azimuthal angle $\phi = 341.186^{\circ}$, giving an extraction radius of $\sim10^{\prime\prime}$ for a region enclosed by 90\% of the PSF. For the background extraction regions we used a circular radius of $40^{\prime\prime}$ at a similar off-axis and azimuthal angle as the source.  The script {\it specextract} was used to extract source and background spectra and generate the response files. Prior to analysis these spectra were again binned at 20 counts per bin to improve statistics. Using the script {\it dmextract} we created a background subtracted lightcurve with a maximum time resolution of 3.2s, before barycentrically correcting this lightcurve using the tool {\it axbary}.

\begin{figure*}
	\includegraphics[angle=-90,width=9cm]{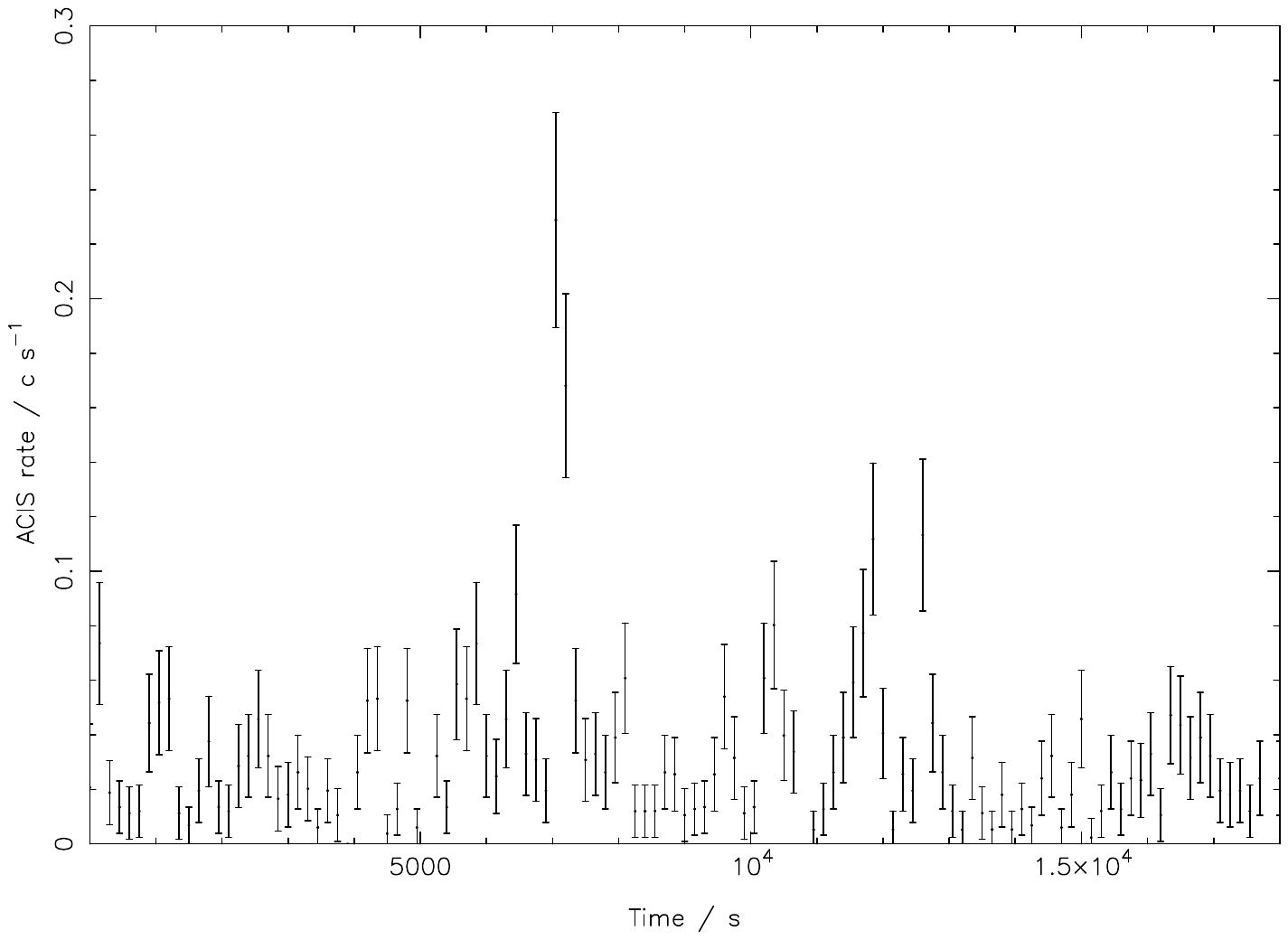}
    \includegraphics[angle=-90,width=9cm]{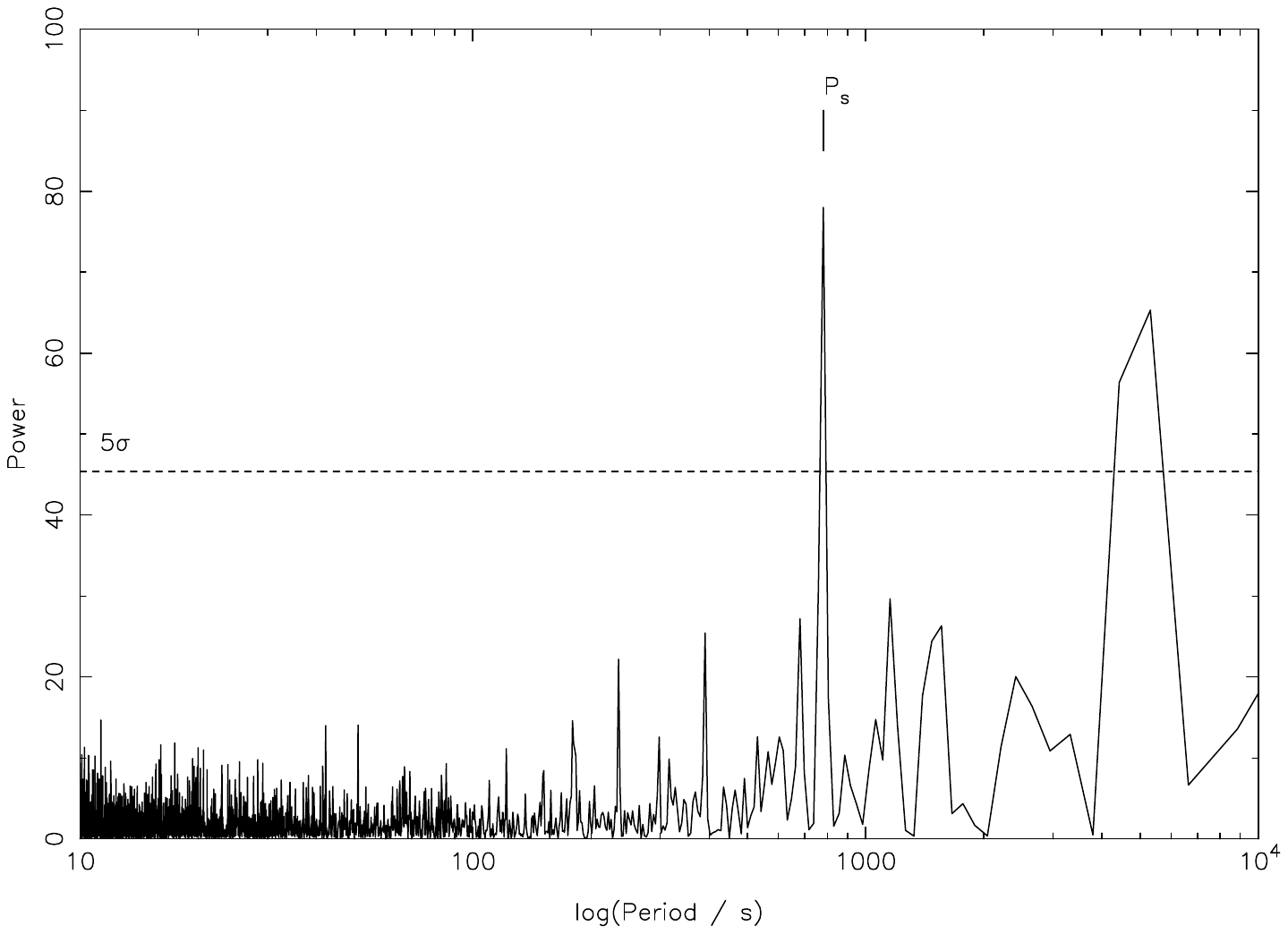}
	\caption{Left: {\it Chandra} background subtracted light curve binned at 150~s. The time axis begins at the start of the observation, at 53197.510 MJD. Right: Power spectrum of the {\it Chandra} background subtracted light curve, showing the detection of modulation at a period of 781~s. The $5 \sigma$ signal detection threshold is indicated by the dashed line.}
	\label{chandra_lc_and_powspec}
\end{figure*}

\section{Timing analysis}
\label{timing}

\subsection{Chandra}
\label{chandra_timing}

Figure 2 displays the background subtracted lightcurve of CXOU~J182531.4$-$144036. It is immediately apparent from this lightcurve that this source exhibits significant variability. The source was previously discussed by \citet{muno08} who detected a 780~s periodic signal in the same archival {\it Chandra} data, when searching for magnetars.

To investigate this further, a power spectrum (Figure 2) was generated for the 3.2~s resolution ACIS-I lightcurve using the Xronos tool {\it powspec}. Upon examination it is apparent that there is a prominent peak that occurs at $\sim 780$~s and a weaker harmonic at half this period. The $5 \sigma$ period detection threshold is indicated on the plot. To further constrain the value of the period of this source, we employed an epoch-folding period search method \citep{larsson96}. Using the HEASOFT task {\it efsearch} we searched for periods between 400 -- 1200 s. The resolution of the period search is defined by the Fourier Period Resolution (FPR) parameter $\delta P = P^{2}/2T_{\text{obs}}$. With $P = 780$~s and $T_{\text{obs}} = 18$~ks, this gives $\delta P \sim 16.9$~s. To improve the likelihood of finding the best period we overestimated the FPR by a factor of $\sim 20$ giving $\delta P = 0.85$~s. We then used this reduced $\delta P$ as the period resolution of the epoch folding period search. The plot of $\chi^{2}$ versus period produced by the {\it efsearch} task shows a broad peak and fitting this with a Gaussian yielded a central value of $781$~s. The uncertainty in the Gaussian central value is $\sim 0.15$~s, so the 1-$\sigma$ confidence level in the period is estimated as this uncertainty multiplied by the FPR overestimation factor, yielding $\pm 3$~s. This period of $781 \pm 3$~s is in agreement with that found by \citet{muno08}. The 0.2 -- 8~keV light curve was then phase folded in 10 phase bins over this best-fitted period and the resulting folded light curve is shown in Figure 3.

\begin{figure}
	\includegraphics[angle=-90,width=\hsize]{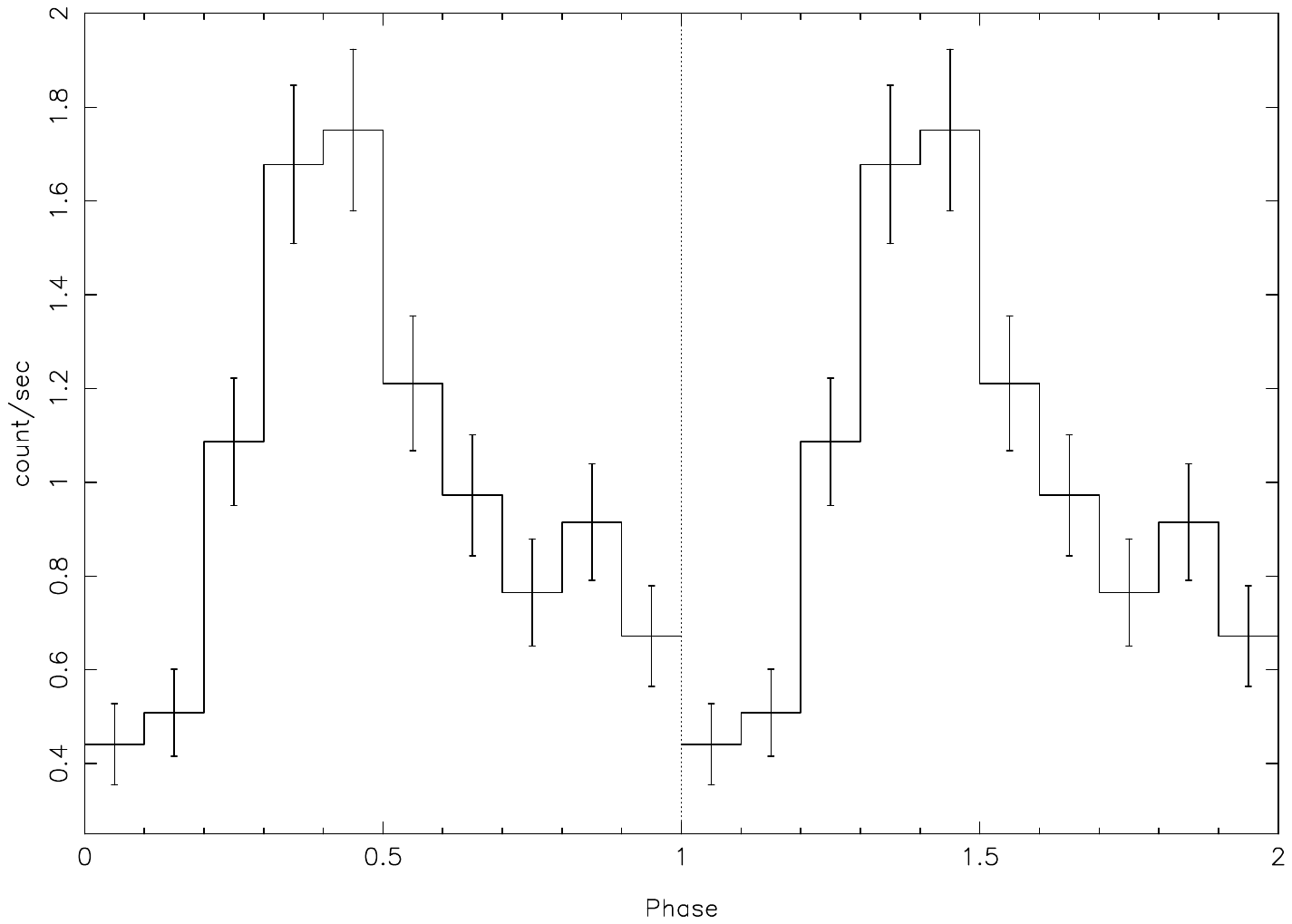}
	\caption{{\it Chandra} background subtracted light curve folded over the best fit period of 781~s. The zero phase is set at the point of minimum observed modulation. The error bars represent the 1-$\sigma$ confidence level.}
	\label{chandra_folded}
\end{figure}

\subsection{XMM-Newton}

The background subtracted and combined EPIC light curve of 4XMM~J182531.5$-$144036 is shown in Figure 4 where it has been binned at 150~s for clarity. As with the {\it Chandra} light curve, it is apparent from Figure 4 that this source displays a noticeable degree of variability. This combined light curve was used to construct a power spectrum, in order to determine the nature of any periodic signal from the source. Figure 4 shows the power spectrum obtained and once again the $5\sigma$ period detection threshold is indicated. The power spectrum obtained is very similar to that from {\it Chandra} in exhibiting a significant peak at $\sim 780$~s.

Using {\it efsearch} we searched for a period in the same time frame as that of the {\it Chandra} source, namely between $400 - 1200$~s. Again, fitting the peak value found from the folded period search with a Gaussian, a central value of 781~s was obtained. The 1-$\sigma$ uncertainty in the period was also obtained using the method described in Section 3.1. The best fitting period was found to be $781 \pm 2$~s. Figure 5 shows the EPIC 0.2 -- 10 keV light curve phase folded at this period.

\begin{figure*}
	\includegraphics[angle=-90,width=9cm]{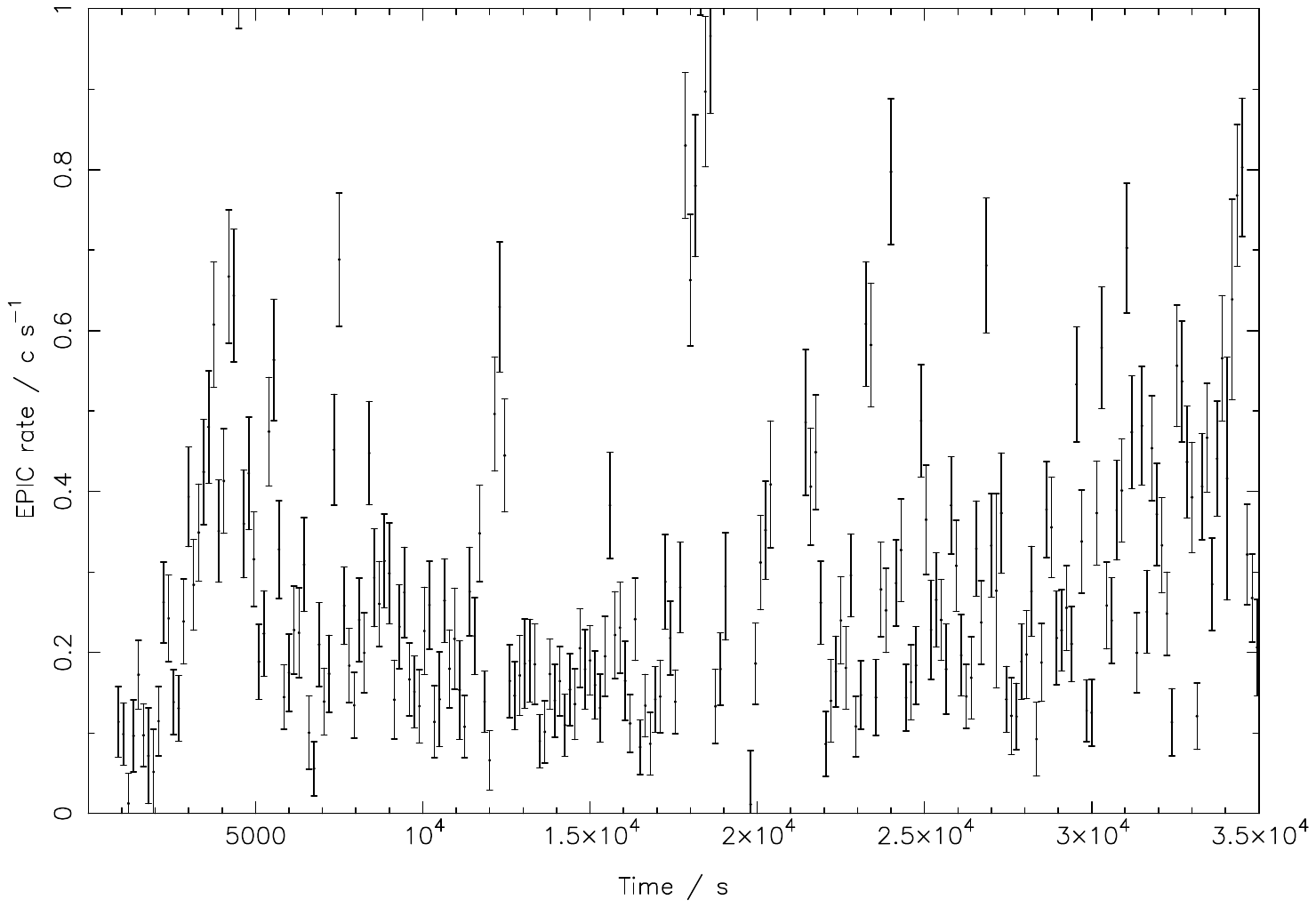}
	\includegraphics[angle=-90,width=9cm]{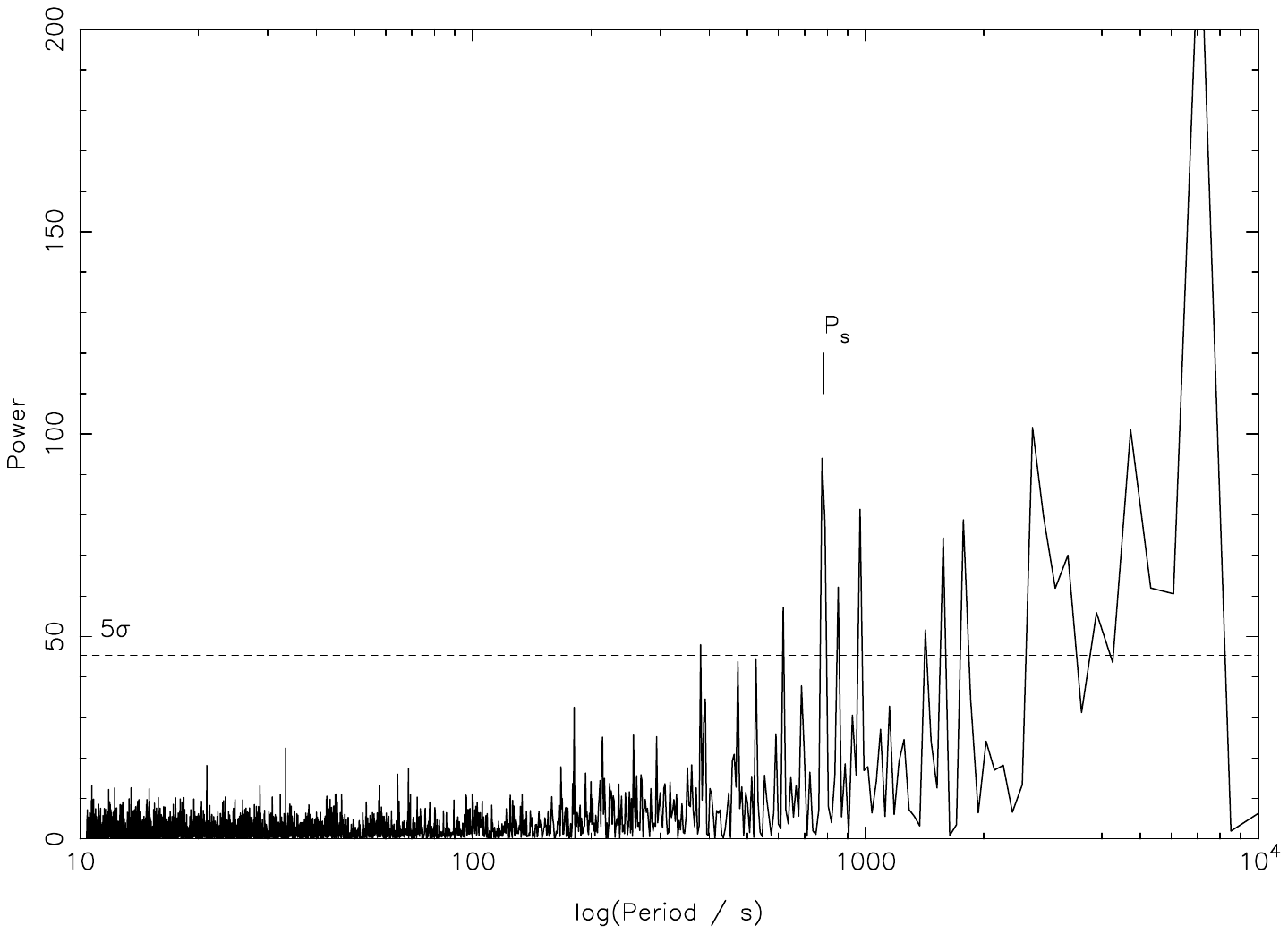}
	\caption{Left: EPIC (PN, MOS1 and MOS2 combined) 0.2 -- 10.0 keV background subtracted light curve binned at 150~s. The time axis indicates the time since the start of the observation at MJD~54566.761. Gaps in the light curve are the result of removing background flares. Right: Power spectrum of the EPIC background subtracted light curve. Below 1000~s the most prominent peak in the spectrum occurs at $\sim 780$~s. The $5 \sigma$ signal detection threshold is indicated by the dashed line.}
	\label{lc_and_powspec}
\end{figure*}

\begin{figure}
	\includegraphics[angle=-90,width=\hsize]{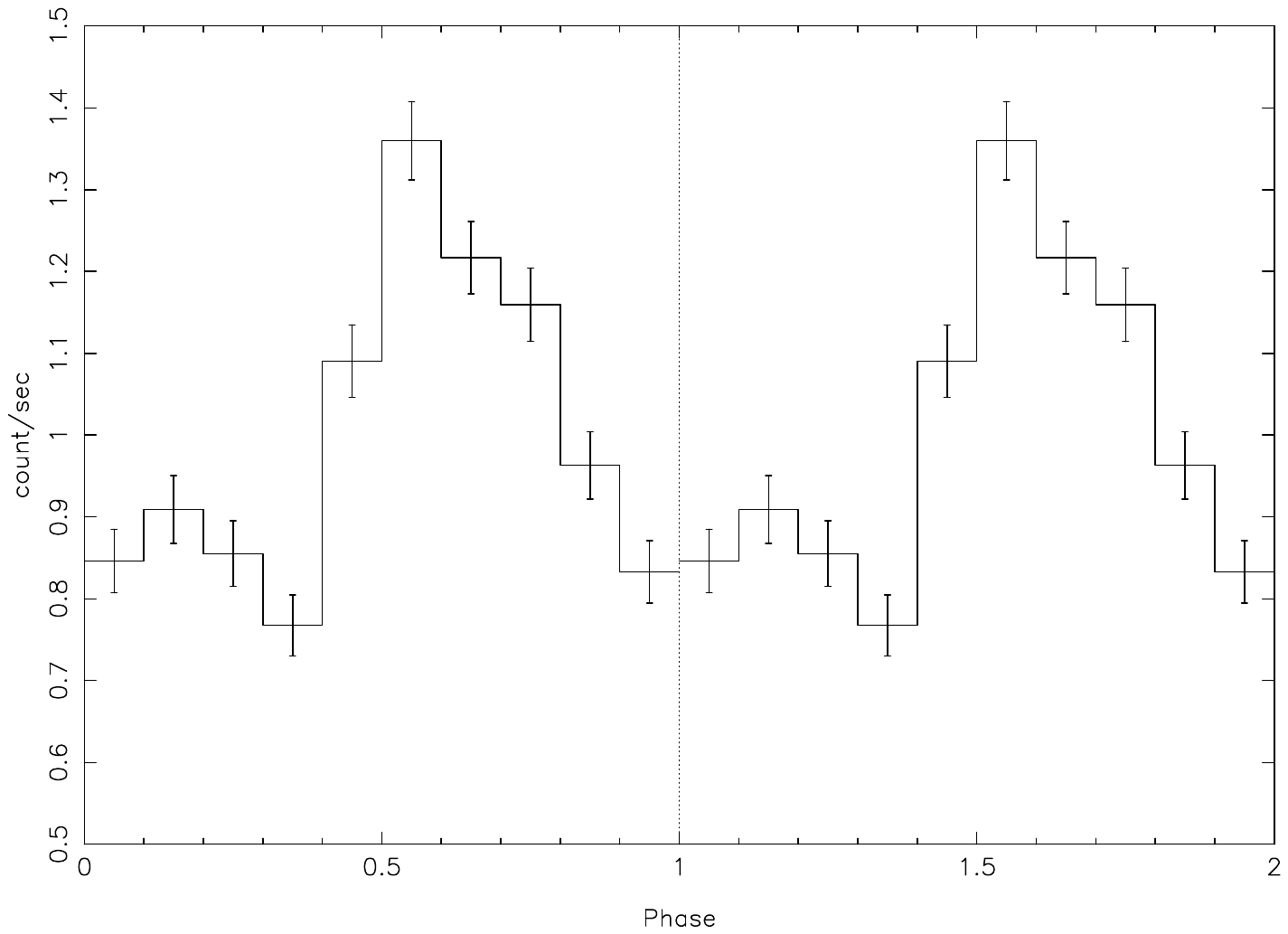}
	\caption{The EPIC 0.2 -- 10 keV background subtracted light curve folded over the best fit period. The zero phase is set at the point of minimum observed modulation. 1-$\sigma$ error bars are shown.}
	\label{folded}
\end{figure}

\section{X-ray Spectral analysis}
\label{spectral}

We extracted energy spectra for 4XMM~J182531.5$-$144036 for all three EPIC cameras in the \textit{XMM-Newton} observation and the ACIS-I camera from the \textit{Chandra} observation. The spectral analysis was performed with XSPEC v.12.8.0 \citep{arnaud96}. The four spectra were fitted simultaneously in the energy range 0.4 -- 10.0 keV with an absorbed power-law, with the hydrogen column density $N_{\text{H}}$, photon index $\Gamma$ and normalisation left as free parameters. Fitting was performed using the C-stat statistic \citep{kaastra17} which is appropriate for spectra with few counts per channel. The absorbed power-law produced an acceptable fit and is shown in Figure 6. The best-fit parameters are $N_{\text{H}} = (6.9 \pm 0.5) \times 10^{22}~\text{atoms~cm}^{-2}$ and  $\Gamma = 2.2 \pm 0.2$. There is no evidence for an iron line at 6.4~keV. The 0.2 -- 10.0 keV unabsorbed flux from the fit is $F_{X} = (7.2 \pm 0.1) \times 10^{-12}$ erg s$^{-1}$ cm$^{2}$. For the regretably unconstrained limits in distance for the source (derived from NIR photometry and the spectral classification of the counterpart, see Section 5.1) of $\sim 1$~kpc and $\sim 7$~kpc, this corresponds to X-ray luminosities of $8 \times 10^{32}$ erg s$^{-1}$ and $4 \times 10^{34}$ erg s$^{-1}$ respectively. We also attempted to fit the spectrum with an exponentially cutoff power law model, which is often seen in Be/X-ray binaries. However this does not improve the fit and the resulting parameter uncertainties are even less well constrained than for the simple power law model.

\begin{figure}[h]
	\includegraphics[width=\hsize]{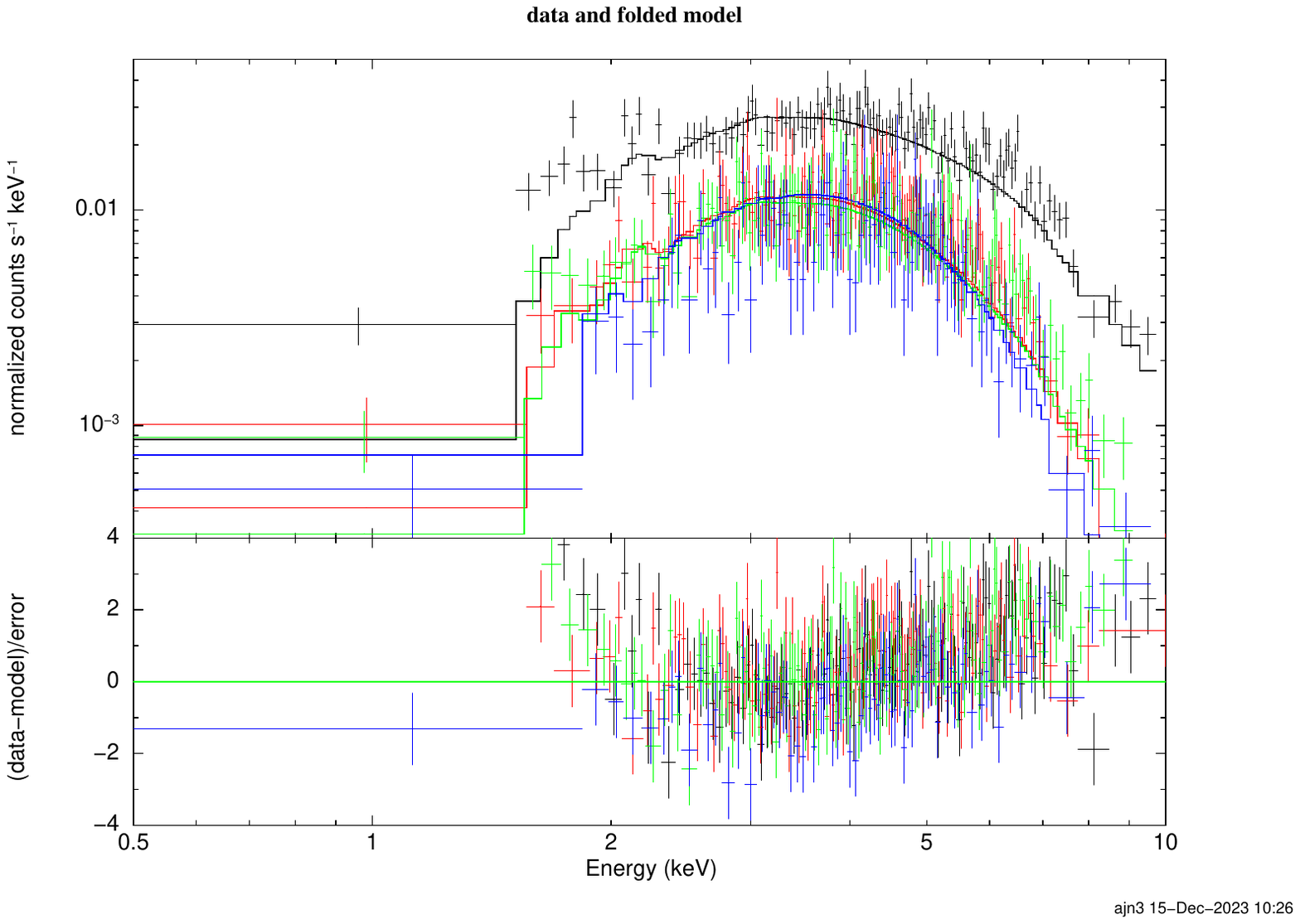}
	\caption{The combined {\it XMM-Newton} EPIC and \textit{Chandra} ACIS-I spectrum of 4XMM~J182531.5$-$144036. (top) The EPIC-PN (black), EPIC-MOS1 (red), EPIC-MOS2 (green) and ACIS-I (blue) data points together with the fitted absorbed power law model. (bottom) The residuals between the data and the best-fit model.}
	\label{spectrum}
\end{figure}

\section{Near infrared observations}
\label{NIR_section}

\subsection{UKIDSS photometry}

\begin{figure}[h]
	\includegraphics[width=\hsize]{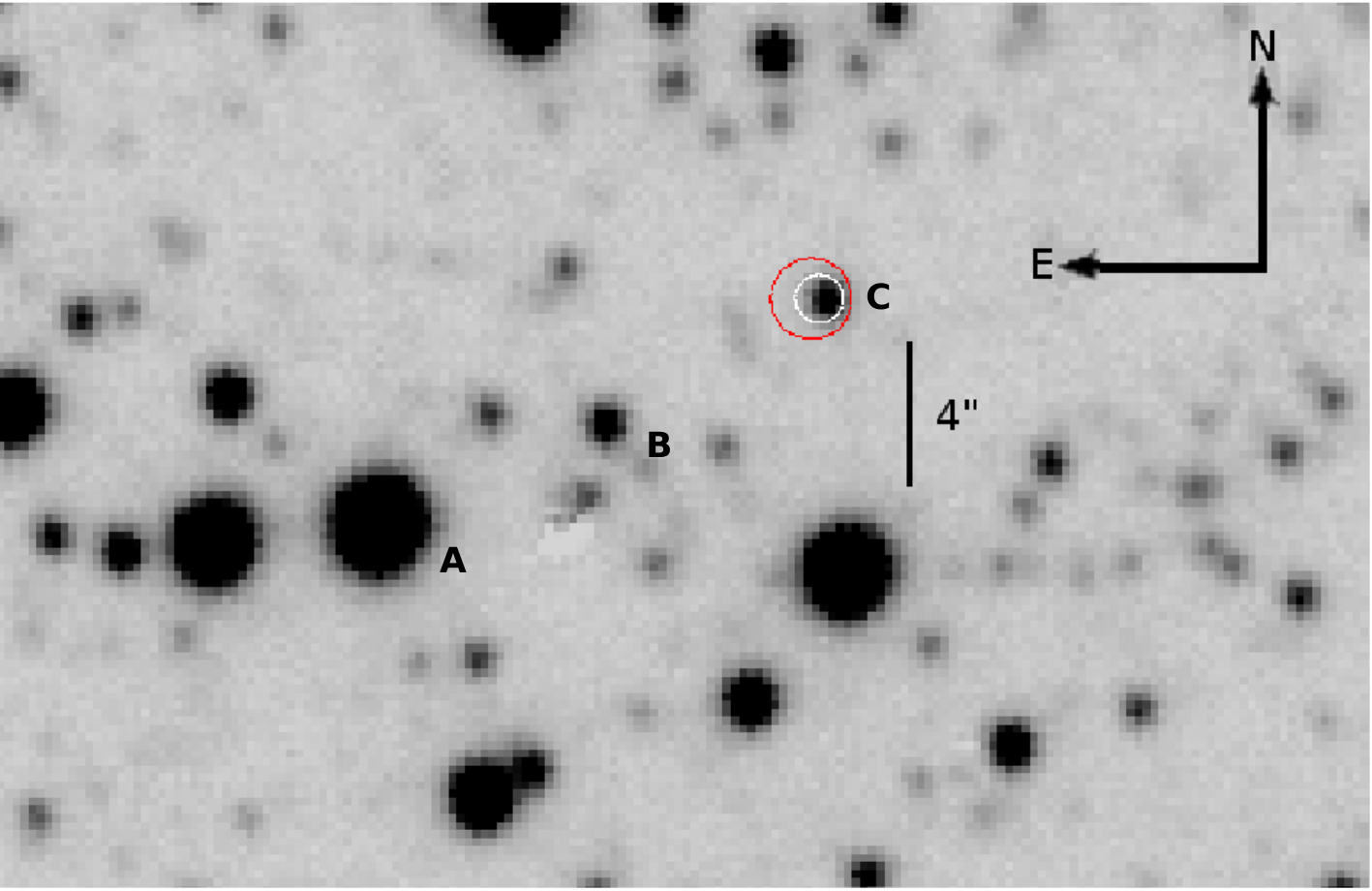}
	\caption{A $25^{\prime \prime} \times  38^{\prime \prime}$ UKIDSS J-band finding chart for 4XMM~ J182531.5$-$144036. The red circle is centred on the {\it XMM-Newton} detected position, with a radius of $1^{\prime \prime}$ equal to the positional error. The white circle is centred on the {\it Chandra} detected position and has a radius of $0.6^{\prime \prime}$ equal to its positional error.}
	\label{finder_chart}
\end{figure}

To search for any possible near infrared (NIR) counterparts to the X-ray source 4XMM~J182531.5$-$144036 we employed the {\it UKIDSS} (UKIRT Infrared Deep Sky Survey) Galactic Plane Survey DR6 photometric data  \citep{warren07} to search a region of 6$^{\prime\prime}$ surrounding its position. There was only one candidate with UKIDSS UGPS~J182531.48--144036.5 located at 0.6$^{\prime\prime}$ from the reported X-ray position. It has infrared magnitudes of $J=16.28$, $H=14.92$ and $K=13.99$ respectively.

Figure 7 shows a J-band {\it UKIDSS} image of the region surrounding the position of 4XMM~J182531.5$-$144036. The white circle represents the {\it Chandra} detected position with a 95$\%$ confidence limit for CXOU J182531.4-144036. The red circle describes the 1-{$\sigma$} positional uncertainty of the source as detected by {\it XMM-Newton}. The NIR counterpart is labelled as object C.

The X-ray spectral fitting gave an equivalent hydrogen column density to the source of $N_{\text{H}} = 6.9 \times 10^{22}$~atoms~cm$^{-2}$. Assuming this to be due to interstellar material, using the relationship of \citet{Predehl}, this implies an optical extinction of $A_V = 38.5$~mag, or in the K-band $A_K = 4.3$~mag ~\citep{Rieke}. Therefore the de-reddened K-band magnitude of the counterpart may be estimated as $K=9.7$ mag. For an assumed spectral type of O9 -- B3 III -- V, the absolute K-band magnitude is in the range $-4.48$ to $-0.25$ \citep{BlumConti} \citep{MartinsPlez}. So the distance may be estimated as between $\sim 1$~kpc and $\sim 7$~kpc. The object is too faint to be detected by \emph{Gaia} so we have no independent distance estimate.

Finally, the {\it UKIDSS} JHK magnitudes of the counterpart are plotted in Figure 8 along with Kurucz model atmosphere spectral energy distributions of O9V, B1V and B3III stars in the range 1.0~$\mu$m to 2.2~$\mu$m. These are normalised to the J-band magnitude of the counterpart, and show that the infrared spectrum of the candidate at wavelengths longer than 1.5~$\mu$m is far brighter than these comparison stellar types.

\begin{figure}
	\includegraphics[width=\hsize]{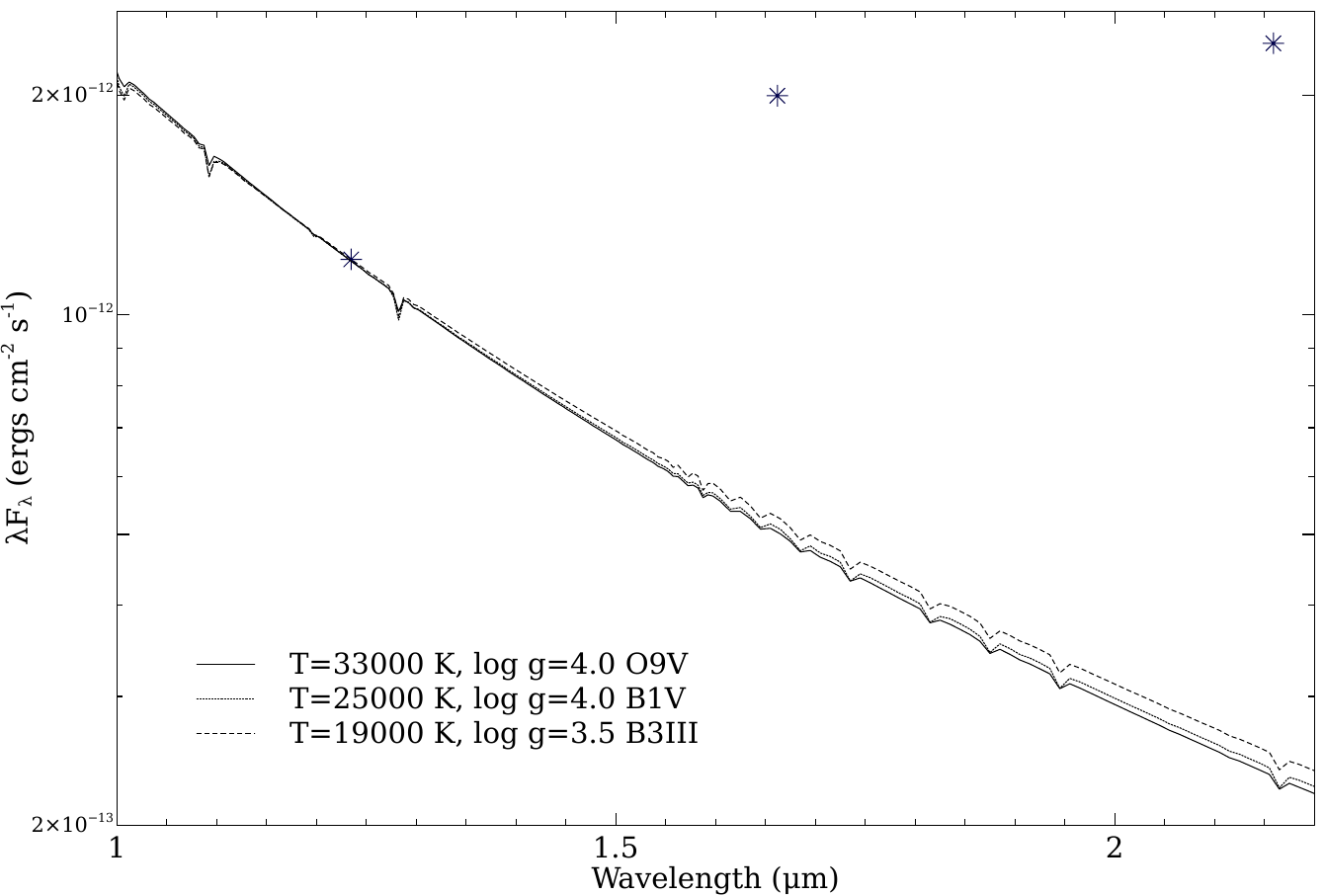}
	\caption{The spectral energy distribution (SED) in the JHK region of the infrared counterpart to 4XMM~J182531.5$-$144036, shown by the three asterisks. Models for the SEDs of O9V, B1V and B3III stars are shown for comparison.}
	\label{SED}
\end{figure}

\subsection{ISAAC spectrum}

The NIR candidate was subsequently observed with the NIR spectrograph {\it ISAAC} on the VLT on 14th May 2011 as part of a programme of follow-up infrared spectra of counterparts to 24 X-ray sources. Due to the faintness of the targets we utilised the SW LRes mode to obtain low-resolution ($R \sim 500$) spectra in the K$_{s}$ band using a $0.8^{\prime\prime}$ wide slit. Nodded science spectra were obtained at a central wavelength of 2.2~$\mu$m for an integration time of 2160~s, followed by observations of the bright (K$_{s}$ = 8.5) B8V telluric standard Hip~092285 for an integration time of 10~s.

The  $120^{\prime\prime}$ long slit was orientated $130^{\circ}$ anti-clockwise from North. As such two further stars, marked A and B in Figure 7 with magnitudes J = 13.23 and J = 16.07 respectively, were also within the slit. The automatic ISAAC pipeline reduction process defaults to extracting the brightest spectrum in the slit, which in this case was star A (whose spectrum included multiple CO bandheads indicative of a red giant). The spectrum of the candidate counterpart C was subsequently extracted manually, and is shown in Figure 9. It displays strong emission lines corresponding to HeI (2.06~$\mu$m) and hydrogen Br$\gamma$ (2.165~$\mu$m). The equivalent width of the Br$\gamma$ line is $7.0 \pm 1.3$~\AA.

\begin{figure}
	\includegraphics[width=\hsize]{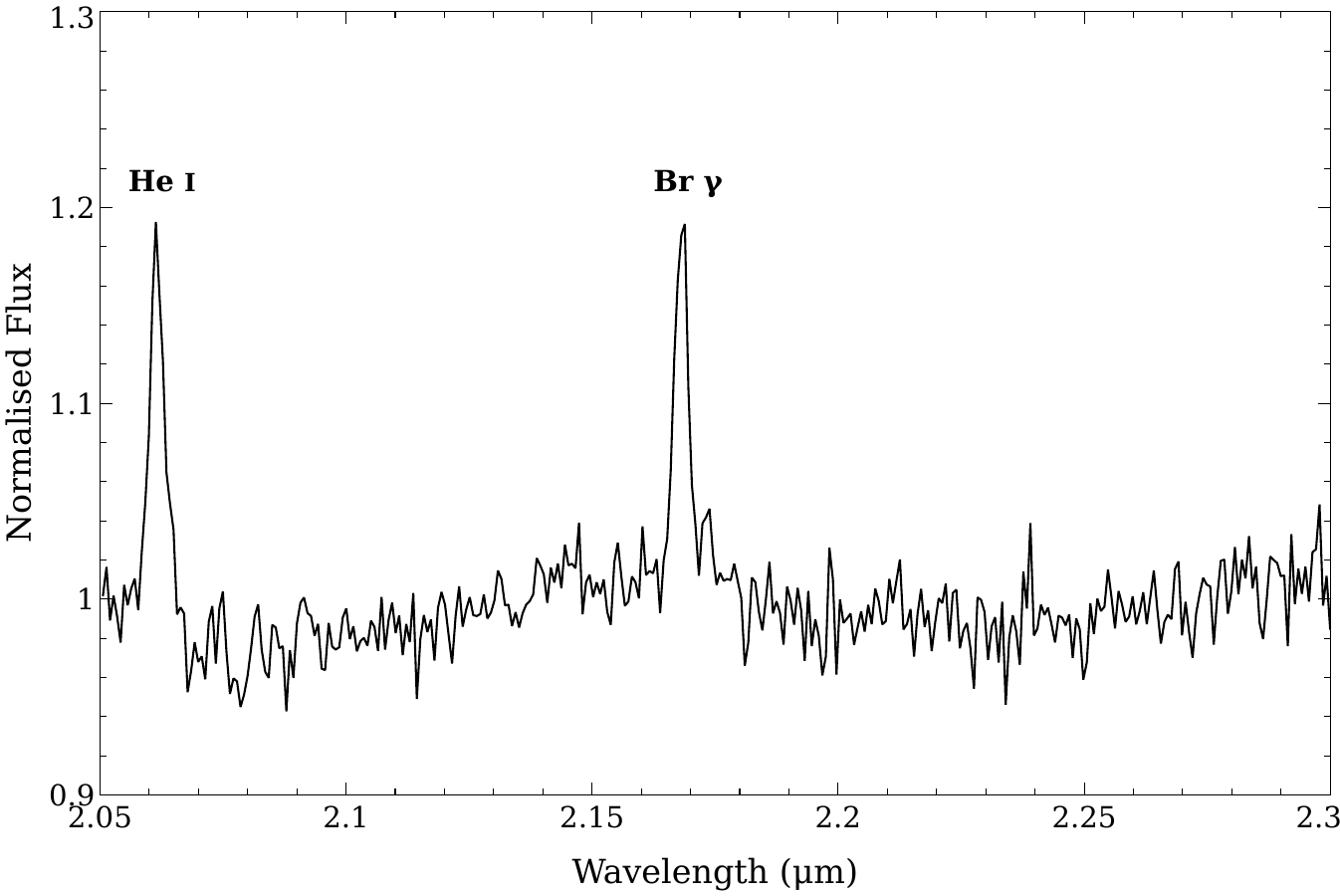}
	\caption{NIR spectrum of the counterpart to 4XMM~J182531.5$-$144036 obtained with {\it ISAAC} on the VLT.}
	\label{NIR_spectrum}
\end{figure}

\section{Discussion}
\label{discussion}

The infrared counterpart to 4XMM~J182531.5$-$144036 displays a near-infrared excess when compared to the spectra of early B-type dwarf or giant stars and exhibits a strong hydrogen emission line. These features are characteristic of the Be star phenomenon. Furthermore, the presence of a coherent X-ray pulsation with a period of 781~s in both the {\it XMM-Newton} and {\it Chandra} data of 4XMM~J182531.5$-$144036 is typical of that seen in BeXRB pulsars \citep{Reig11} while the asymmetric nature of the X-ray pulse profile is also typical of accreting X-ray pulsars and can provide information about the magnetic field structure of the neutron star \citep{meszaros84}. The fact that the X-ray pulsation is seen with the same profile in widely separated observations ({\it XMM-Newton} from April 2008 and {\it Chandra} from July 2004) indicates that the X-ray emission is likely persistent. The hard X-ray spectrum, lack of an iron line, and (poorly constrained) inferred luminosity are also consistent with other wide-orbit, low eccentricity BeXRBs with relatively slow neutron star spin periods ($> 200$~s), such as X Per \citep{ReigRoche}. Considering the locus of BeXRB systems on the Corbet diagram \citep{Reig11}, we may estimate that this object has an orbital period in the range $\sim 250 - 500$~d.  We therefore conclude that 4XMM~J182531.5$-$144036 is a newly identified persistent, long period, Be/X-ray binary.

\begin{acknowledgement}
This work is based on data obtained as part of the UKIRT Infrared Deep Sky Survey and on observations collected at the European Organisation for Astronomical Research in the Southern Hemisphere under ESO programme 087.D-0873(A). It also uses observations obtained with \emph{XMM-Newton}, an ESA science mission with instruments and contributions directly funded by ESA Member States and NASA, and data obtained from the \emph{Chandra} Data Archive and the \emph{Chandra} Source Catalog, with software provided by the \emph{Chandra} X-ray Center (CXC) in the application packages CIAO and Sherpa. We thank the anonymous referee for several useful suggestions.
\end{acknowledgement}

\printbibliography
\end{document}